\documentclass[12pt]{article}
\usepackage{numprint}
\usepackage{amsmath}
\usepackage{graphicx}
\usepackage{tabularx}
\usepackage{subfigure}
\usepackage{float}
\usepackage{longtable}
\usepackage{color}
\usepackage{times,natbib}
\usepackage{amsfonts}
\usepackage{verbatim}
\usepackage{fixltx2e}
\usepackage[labelfont=bf]{caption}
\usepackage[hyperindex,breaklinks]{hyperref}
\usepackage{breakurl}

\title{Fast Bayesian parameter estimation for stochastic logistic growth models}
\author{Jonathan Heydari \and Conor Lawless \and David A. Lydall \and Darren J. Wilkinson\\[1ex] Newcastle University, UK}

\begin{document}
 \maketitle
\begin{abstract}
The transition density of a stochastic, logistic population growth model with multiplicative intrinsic noise is analytically intractable.
Inferring model parameter values by fitting such stochastic differential equation (SDE) models to data therefore requires relatively slow numerical simulation.
Where such simulation is prohibitively slow, an alternative is to use model approximations which do have an analytically tractable transition density, enabling fast inference.
We introduce two such approximations, with either multiplicative or additive intrinsic noise, each derived from the linear noise approximation of the logistic growth SDE. 
After Bayesian inference we find that our fast LNA models, using Kalman filter recursion for computation of marginal likelihoods, give similar posterior distributions to slow arbitrarily exact models.
We also demonstrate that simulations from our LNA models better describe the characteristics of the stochastic logistic growth models than a related approach.
Finally, we demonstrate that our LNA model with additive intrinsic noise and measurement error best describes an example set of longitudinal observations of microbial population size taken from a typical, genome-wide screening experiment.

\end{abstract}

\section{\label{sec:Introduction}Introduction}

Stochastic models simultaneously describe dynamics and noise or heterogeneity in real systems \citep{calibayes}.  For example, stochastic models are increasingly recognised as necessary tools for understanding the behaviour of complex biological systems \citep{wilkinson2012stochastic,wilkinson_nature} and are also used to capture uncertainty in financial market behaviour \citep{stochfinance,stochfinance2}.  Many such models are written as continuous stochastic differential equations (SDEs) which often do not have analytical solutions and are slow to evaluate numerically compared to their deterministic counterparts.  Simulation speed is often a particularly critical issue when inferring model parameter values by comparing simulated output with observed data \citep{woodtrees}.  

For SDE models where no explicit expression for the transition density is available, it is possible to infer parameter values by simulating a latent process using a data augmentation approach \citep{darren2005}. However, this method is computationally intensive and not practical for all applications. When fast inference for SDEs is important, for example for real-time analysis as part of decision support systems or for big data inference problems where we simultaneously fit models to many thousands of datasets (e.g. \cite{heydari}), we need an alternative approach.
Here we demonstrate one such approach: developing an analytically tractable approximation to the original SDE, by making linear noise approximations (LNAs).  We apply this approach to a SDE describing logistic population growth for the first time.
  
The logistic model of population growth, an ordinary differential equation (ODE) describing the self-limiting growth of a population of size $X_t$ at time $t$, was developed by \cite{Verhulst1847}
\begin{align}
\label{eq_det}
\frac{dX_t}{dt}&=rX_t-\frac{r}{K}X_t^2.
\end{align}
The ODE has the following analytic solution:
\begin{align}
\label{eq_det_sol}
X_t&=\frac{K}{1+Qe^{-r (t-t_0)}},
\end{align}
where $Q=\left(\frac{K}{P}-1\right)e^{rt_0}$ and $P=X_{t_0}$.
The model describes a population growing from an initial size $P$ with an intrinsic growth rate $r$, undergoing approximately exponential growth which slows as the availability of some critical resource (e.g. nutrients or space) becomes limiting \citep{theoryoflogisticgro}.
Ultimately, population density saturates at the carrying capacity (maximum achievable population density) $K$, once the critical resource is exhausted.
Where further flexibility is required, generalized forms of the logistic growth process \citep{analysisoflogistic,logisticrevisited} may be used instead.

To account for uncertainty about processes affecting population growth which are not explicitly described by the deterministic logistic model, we can include a term describing intrinsic noise and consider a SDE version of the model \citep{optimal_harvesting,roman}.  Here we extend the ODE in (\ref{eq_det}) by adding a term representing multiplicative intrinsic noise (\ref{eq_det_sde}) to give a model which we refer to as the stochastic logistic growth model (SLGM).  
\begin{align}
\label{eq_det_sde}
dX_t&=\left[rX_t-\frac{r}{K}X_t^2\right]dt+\sigma X_t dW_t, 
\end{align}
where $X_{t_0}=P$ and is independent of $W_t$, $t\geq t_0$, 

Alternative stochastic formulations of the logistic ODE can be generated \citep{stolog}.
The Kolmogorov forward equation has not been solved for (\ref{eq_det_sde}) (or for any similar formulation of a logistic SDE) and so no explicit expression for the transition density is available.

\cite{roman} introduce a diffusion process approximating the SLGM (which we label RRTR) with a transition density that can be derived explicitly.  
The Bayesian approach can be applied in a natural way to carry out parameter inference for state space models with tractable transition densities \citep{dynamicmodels}.  
A state space model describes the probabilistic dependence between an observation process variable $X_t$ and state process $S_t$.
The transition density is used to describe the state process $S_t$ and a measurement error structure is chosen to describe the relationship between $X_t$ and $S_t$. 

The Kalman filter \citep{kalmanoriginal} is typically used to infer the hidden state process of interest $S_t$ and is an optimal estimator, minimising the mean square error of estimated parameters when all noise in the system can be assumed to be Gaussian.
We use the Kalman filter to reduce computational time in a parameter inference algorithm by recursively computing the marginal likelihood \citep{dynamicmodels}.

The RRTR can be fit to data within an acceptable time frame by assuming multiplicative measurement error to give a linear Gaussian structure, allowing us to use a Kalman filter for inference.

We introduce two new first order linear noise approximations (LNAs) \citep{LNA,komorowski} of (\ref{eq_det_sde}), one with multiplicative and one with additive intrinsic noise, which we label LNAM and LNAA respectively.
The LNA reduces a SDE to a linear SDE with additive noise, which can be solved to give an explicit expression for the transition density. 
We derive transition densities for the two approximate models and construct a Kalman filter by choosing measurement noise to be either multiplicative or additive to construct a linear Gaussian structure. 
Exact simulations from the SGLM are compared with each of the three approximate models.
We compare the utility of each of the approximate models during parameter inference by comparing simulations with both synthetic and real datasets.
 
\section{\label{sec:Roman}The \cite{roman} diffusion process}

\cite{roman} present a logistic growth diffusion process (RRTR) which has a transition density that can be written explicitly, allowing inference of model parameter values from discrete sampling trajectories.
 \\
The RRTR is derived from the following ODE:
\begin{align}
\label{eq_ode}
\frac{dx_t}{dt}&=\frac{Qr}{e^{rt}+Q}x_t.
\end{align}
The solution to (\ref{eq_ode}) is given in (\ref{eq_det_sol}) (it has the same solution as (\ref{eq_det})).

\cite{roman} see (\ref{eq_ode}) as a generalisation of the Malthusian growth model with a deterministic, time-dependent fertility $h(t)=\frac{Qr}{e^{rt}+Q}$, and replace this with $\frac{Qr}{e^{rt}+Q}+\sigma W_t$ to obtain the following approximation to the SLGM:
\begin{align}
\label{eq_sde}
dX_t&=\frac{Qr}{e^{r{t}}+Q}X_td{t}+{\sigma}X_tdW_t,
\end{align}
where  $Q=\left(\frac{K}{P}-1\right)e^{rt_0}$, $P=X_{t_0}$ and is independent of $W_t$, $t\geq t_0$.
The process described in (\ref{eq_sde}) is a particular case of the lognomal process with exogenous factors, therefore an exact transition density is available \citep{gutierrez}.
The transition density for $Y_t$, where $Y_t=\log(X_t)$, can be written:
\begin{align}
\begin{split}\label{eq:RRTR_tran}
(Y_{t_i}|Y_{t_{i-1}} &= y_{t_{i-1}}) \sim\operatorname{N}\left(
\mu_{t_i}
,
\Xi_{t_i}
\right),\\
\text{where } a &= r,\quad b=\frac{r}{K}, \\
\mu_{t_i}=& \log(y_{t_{i-1}})
+\log(\frac{1+be^{-at_i}}{1+be^{-at_{i-1}}})
-\frac{\sigma^2}{2}(t_i-t_{i-1}) \text{ and}\\
\Xi_{t_i} &= \sigma^2(t_i-t_{i-1}).
\end{split}
\end{align}
 
\section{\label{sec:LNAM} Linear noise approximation with multiplicative noise}
We now take a different approach to approximating the SLGM (\ref{eq_det_sde}), which will turn out to be closer to the exact solution of the SLGM than the RRTR (\ref{eq_sde}). 
Starting from the original model (\ref{eq_det_sde}),
we apply It\^{o}'s lemma with the transformation $f(t,X_t)\equiv Y_t=\log X_t$
to obtain the following It\^{o} drift-diffusion process:
\begin{align}
\label{eq:SDE2}
dY_t=\left(r-\frac{1}{2}\sigma^2-\frac{r}{K}e^{Y_t}\right)dt+\sigma dW_t.
\end{align}
The log transformation from multiplicative to additive noise, gives a constant diffusion term, so that the LNA will give a good approximation to (\ref{eq_det_sde}).  
We now separate the process $Y_t$ into a deterministic part $V_t$ and a stochastic part $Z_t$ so that $Y_t=V_t+Z_t$ and consequently $dY_t=dV_t+dZ_t$.
We choose $V_t$ to be the solution of the deterministic part
\begin{equation*}
dV_t=\left(r-\frac{1}{2}\sigma^2-\frac{r}{K}e^{V_t}\right)dt.
\end{equation*}
After redefining our notation as follows: $a=r-\frac{\sigma^2}{2}$ and $b=\frac{r}{K}$, we solve (\ref{eq:SDE2}) for $V_t$:
\begin{equation}\label{eq:LNAM_det_sol}
V_t=\log\left(\frac{aPe^{aT}}{bP(e^{aT}-1)+a}\right).
\end{equation}
Differentiating w.r.t. t, we obtain: 
\begin{equation*} \label{eq:LNAM_dv_t}
\frac{dV_t}{dt}=\frac{a(a-bP)}{bP(e^{aT}-1)+a}.
\end{equation*}
Writing down an expression for $dZ_t$, where $dZ_t=dY_t-dV_t$.
\begin{equation*} 
dZ_t=\left(r-\frac{1}{2}\sigma^2-\frac{r}{K}e^{Y_t}\right)dt+\sigma dW_t-\frac{a(a-bP)}{bP(e^{aT}-1)+a}dt
\end{equation*}
and simplifying:
\begin{equation*}
dZ_t=\left(\frac{baPe^{aT}}{bP(e^{aT}-1)+a}-be^{Y_t}\right)dt+\sigma dW_t.
\end{equation*}
Recognizing that $e^{V_t}=\frac{aPe^{aT}}{bP(e^{aT}-1)+a}$:
\begin{equation*} 
dZ_t=b\left(e^{V_t}-e^{Y_t}\right)dt+\sigma dW_t.
\end{equation*}
We substitute in $Y_t=V_t+Z_t$ to give
\begin{equation*} 
dZ_t=b\left(e^{V_t}-e^{V_t+Z_t}\right)dt+\sigma dW_t.
\end{equation*} 
We now apply the LNA, by making a first-order approximation of $e^{Z_t}\approx 1+Z_t$ and then simplify to give $dZ_t=-be^{V_t}Z_tdt+\sigma dW_t$.
Finally, substitute $e^{V_t}=\frac{aPe^{aT}}{bP(e^{aT}-1)+a}$  to obtain
\begin{equation}\label{eq:LNAM_dz}
dZ_t=-\frac{baPe^{aT}}{bP(e^{aT}-1)+a}Z_tdt+\sigma dW_t.
\end{equation}
This process is a particular case of the time-varying Ornstein-Uhlenbeck process, which can be solved explicitly. 
The transition density for $Y_t$ (derivation in Appendix~\ref{app:LNAM_sol}) is then:
\begin{align}
\begin{split}\label{eq:LNAM_tran}
(Y_{t_i}|Y_{t_{i-1}}&=y_{t_{i-1}})\sim\operatorname{N}\left(\mu_{t_i},\Xi_{t_i}\right),\\
\text{redefine } y_{t_{i-1}}&=v_{t_{i-1}}+z_{t_{i-1}}, Q=\left(\frac{\frac{a}{b}}{P}-1\right)e^{at_{0}},\\
\mu_{t_i}&=v_{t_{i-1}}+\log\left(\frac{1+Qe^{-at_{i-1}}}{1+Qe^{-at_i}}\right)+e^{-a(t_i-t_{i-1})}\frac{1+Qe^{-at_{i-1}}}{1+Qe^{-at_i}}z_{t_{i-1}} \text{ and}\\
\Xi_{t_i}&=\sigma^2\left[\frac{4Q(e^{at_i}-e^{at_{i-1}})+e^{2at_i}-e^{2at_{i-1}}+2aQ^2(t_i-t_{i-1})}{2a(Q+e^{at_i})^2}\right].
\end{split}
\end{align}
The LNA of the SLGM with multiplicative intrinsic noise (LNAM) can then be written as
\begin{align*}
d\log X_t=\left[dV_t+be^{V_t}V_t-be^{V_t}\log X_t\right]dt+\sigma dW_t,
\end{align*}
where $P=X_{t_0}$ and is independent of $W_t$, $t\geq t_0$.
\\
Note that the RRTR given in (\ref{eq_sde}) can be similarly derived using a zero-order noise approximation ($e^{Z_t}\approx 1$) instead of the LNA.
 
\section{\label{sec:LNAA} Linear noise approximation with additive noise}
As in Section~\ref{sec:LNAM}, we start from the SLGM, given in (\ref{eq_det_sde}).
Without first log transforming the process, the LNA will lead to a worse approximation to the diffusion term of the SLGM, but we will see in the coming sections that there are nevertheless advantages.
We separate the process $X_t$ into a deterministic part $V_t$ and a stochastic part $Z_t$ so that $dX_t=dV_t+dZ_t$ and consequently $X_t=V_t+Z_t$.
We choose $V_t$ to be the solution of the deterministic part
\begin{equation*}
dV_t=\left(rV_t-\frac{r}{K}V_t^2\right)dt.
\end{equation*}
After redefining our previous notation as follows: $a=r$ and $b=\frac{r}{K}$, we solve $dV_t$ to give:
\begin{equation*}
V_t=\frac{aPe^{aT}}{bP(e^{aT}-1)+a}.
\end{equation*}
Differentiating w.r.t.~$t$, we obtain: 
\begin{equation*}
\frac{dV_t}{dt}=\frac{a^2Pe^{aT}(a-bP)}{\left(bP(e^{aT}-1)+a\right)^2}.
\end{equation*}
We now solve $dZ_t$, where $dZ_t=dX_t-dV_t$. Expressions for both $dX_t$ and $dV_t$ are known:
\begin{equation*}
dZ_t=\left(rX_t-\frac{r}{K}X_t^2\right)dt+\sigma X_t dW_t-\frac{a^2Pe^{aT}(a-bP)}{\left(bP(e^{aT}-1)+a\right)^2}dt.
\end{equation*}
Simplifying:
\begin{equation*}
dZ_t=\left(aX_t-bX_t^2-\frac{a^2Pe^{aT}(a-bP)}{\left(bP(e^{aT}-1)+a\right)^2}\right)dt+\sigma X_t dW_t.
\end{equation*}
We then substitute in $X_t=V_t+Z_t$ and rearrange to give
\begin{align*}
dZ_t=&\left(aV_t-bV_t^2-\frac{a^2Pe^{aT}(a-bP)}{\left(bP(e^{aT}-1)+a\right)^2}+(a-2bV_t)Z_t-bZ_t^2\right)dt\\
&+\left( \sigma V_t +\sigma Z_t\right) dW_t.
\end{align*}
We now apply the LNA, by setting second-order terms $-bZ^2dt=0$ and $\sigma Z_t dW_t=0$ to obtain
\begin{equation}\label{eq:LNAA_dz}
dZ_t=\left(aV_t-bV_t^2-\frac{a^2Pe^{aT}(a-bP)}{\left(bP(e^{aT}-1)+a\right)^2}+(a-2bV_t)Z_t\right)dt+\sigma V_t  dW_t.
\end{equation}
This process is a particular case of the Ornstein-Uhlenbeck process, which can be solved. 
The transition density for $X_t$ (derivation in Appendix~\ref{app:LNAA_sol}) is then
\begin{align}
\begin{split}\label{eq:LNAA_tran}
(X_{t_i}|X_{t_{i-1}}&=x_{t_{i-1}})\sim N(\mu_{t_i},\Xi_{t_i}),\\
\text{where } x_{t_{i-1}}&=v_{t_{i-1}}+z_{t_{i-1}},\\
\mu_{t_i}&=v_{t_{i-1}}+\left(\frac{aPe^{aT_i}}{bP(e^{aT_i}-1)+a}\right)-\left(\frac{aPe^{aT_{i-1}}}{bP(e^{aT_{i-1}}-1)+a}\right)\\
&+e^{a(t_i-t_{i-1})}\left(\frac{bP(e^{aT_{i-1}}-1)+a}{bP(e^{aT_i}-1)+a}\right)^2Z_{t_{i-1}}\text{ and}\\
 \Xi_t&=\frac{1}{2}\sigma^2aP^2e^{2aT_i}\left(\frac{1}{bP(e^{aT_i}-1)+a}\right)^4\\
&\times[
b^2P^2(e^{2aT_i}-e^{2aT_{i-1}})
+4bP(a-bP)(e^{aT_i}-e^{aT_{i-1}})\\
&\;\:\:\:\:+2a(t_i-t_{i-1})(a-bP)^2
].
\end{split}
\end{align}
The LNA of the SLGM, with additive intrinsic noise (LNAA) can then be written as
\begin{align*}
dX_t=\left[b{V_t}^2+\left(a-2bV_t\right)X_t\right]dt+\sigma V_t dW_t,
\end{align*}
where $P=X_{t_0}$ and is independent of $W_t$, $t\geq t_0$.
\section{\label{sec:SDE_application}Simulation and Bayesian inference for logistic SDE and approximations}
\begin{figure}[h!]
  \centering
\includegraphics[width=14cm]{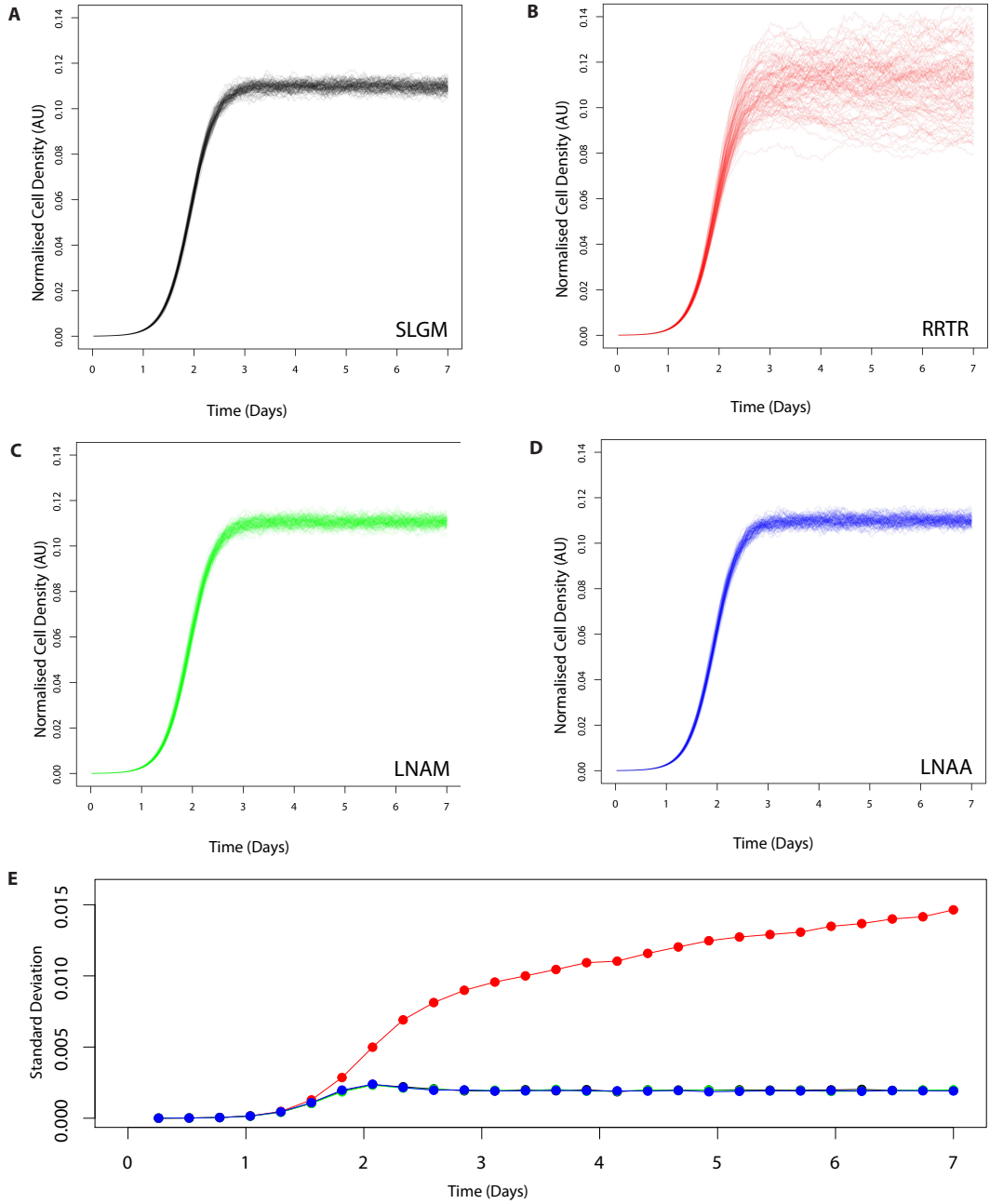}
\caption{\label{4nonu} 
Forward trajectories (No. of simulations=100) for logistic SDE and approximations.
See Table~\ref{app:sde_val_fur} for parameter values. 
A) The stochastic logistic growth model (SLGM). 
B) The \cite{roman} (RRTR) approximation. 
C) The linear noise approximation with multiplicative intrinsic noise (LNAM).
D) The linear noise approximation with additive intrinsic noise (LNAA).
E) Standard deviations of simulated trajectories over time for the SLGM (black), RRTR (red), LNAM (green) and LNAA (blue).
}
\end{figure}

To test which of the three approximate models best represent the SLGM, we first compare simulated forward trajectories from the RRTR, LNAM and LNAA with simulated forward trajectories from the SLGM (Figure~\ref{4nonu}).
We use the Euler-Maruyama method \citep{embook} with very fine discretisation to give arbitrarily exact simulated trajectories from each SDE.

The LNAA and LNAM trajectories are visually indistinguishable from the SLGM (Figure~\ref{4nonu} A,C \& D).
On the other hand, population sizes simulated with the RRTR display large deviations from the mean as the population approaches stationary phase (Figure~\ref{4nonu}A \& B). 
Figure~\ref{4nonu}E further highlights the increases in variation as the population approaches stationary phase for simulated trajectories of the RRTR, in contrast to the SLGM and LNA models.

\subsection{\label{sec:simulation_stu}Bayesian parameter inference with approximate models}
To compare the quality of parameter inference using each of these approximations we simulated synthetic time-course data from the SLGM and combined this with either lognormal or normal measurement error. 
Carrying out Bayesian inference with broad priors (see (\ref{app:LNAM_sta_spa_mod}) and (\ref{app:LNAA_sta_spa_mod})) we compared the parameters recovered using each approximation with those used to generate the synthetic dataset. 
The synthetic time-course datasets consist of 27 time points generated using the Euler-Maruyama method with very fine intervals \citep{euler_maruyama}.

We formulate our inference problem as a dynamic linear state space model.  
To allow fast parameter inference we take advantage of a linear Gaussian structure and construct a Kalman filter recursion for marginal likelihood computation (Appendix~\ref{app:kalman_fil}).
We therefore assume lognormal (multiplicative) error for the RRTR and LNAM, and for the LNAA we assume normal (additive) measurement error.
Dependent variable $y_{t_i}$ and independent variable $\{t_{i},i=1,...,N\}$ are data input to the model (where $t_i$ is the time at point $i$ and $N$ is the number of time points). 
$X_t$ is the state process, describing the population size.
See Table~\ref{table:SDE_priors} for prior hyper-parameter values.

For the RRTR and LNAM,
\begin{align}
\log(y_{t_i}) &\sim \operatorname{N}(X_{t_i},{\nu}^{2} ),\notag\\
(X_{t_i}|X_{t_{i-1}}=x_{t_{i-1}})&\sim\operatorname{N}\left(\mu_{t_i},\Xi_{t_i}\right), \text{ where } x_{t_{i}}=v_{t_{i}}+z_{t_{i}},\label{app:LNAM_sta_spa_mod}
\end{align}
$\mu_{t_i}$ and $\Xi_{t_i}$ are given by (\ref{eq:RRTR_tran}) and (\ref{eq:LNAM_tran}) for the RRTR and LNAM respectively. Priors are as follows:
\begin{align*}
\log X_0 \equiv \log~P &\sim \operatorname{N}({\mu}_P,{\tau_P}^{-1}),& \quad\quad
\log~K &\sim \operatorname{N}({\mu}_K,{\tau_K}^{-1}),& \quad\quad
\log~r &\sim \operatorname{N}({\mu}_r,{\tau_r}^{-1}),\\
\log~\nu^{-2} &\sim \operatorname{N}({\mu}_\nu,{\tau_\nu}^{-1}),&
\log~{\sigma^{-2}} &\sim \operatorname{N}({\mu}_\sigma,{\tau_\sigma}^{-1})I_{[1,\infty]}.&
\end{align*}

For the LNAA, 
\begin{align}
y_{t_i} &\sim \operatorname{N}(X_{t_i},{\nu}^{2} ),\notag\\
(X_{t_i}|X_{t_{i-1}}=x_{t_{i-1}})&\sim\operatorname{N}\left(\mu_{t_i},\Xi_{t_i}\right)
, \text{ where } x_{t_{i}}=v_{t_{i}}+z_{t_{i}},\label{app:LNAA_sta_spa_mod}
\end{align}
$\mu_{t_i}$ and $\Xi_{t_i}$ are given by (\ref{eq:LNAA_tran}). Priors are as in (\ref{app:LNAM_sta_spa_mod}).
Our prior for $\log~{\sigma^{-2}}$ is truncated below 1 to avoid unnecessary exploration of extremely low probability regions, which could be caused when there are problems identifying $\nu$, for example when $\log~\nu^{-2}$ takes large values, and to ensure that intrinsic noise does not dominate the process.
The truncation limit was chosen by visual inspection.

To see how the inference from our approximate models compares with slower ``exact'' models, we consider Euler-Maruyama approximations \citep{euler_maruyama} of (\ref{eq_det_sde}) and of the log transformed process, using fine intervals.
Given these approximations we can construct a state space model for an ``exact'' SLGM with lognormal measurement error (SLGM+L) and similarly for the SLGM with normal measurement error (SLGM+N), priors are as in (\ref{app:LNAM_sta_spa_mod}).

Model fitting is carried out using standard MCMC techniques (the Gibbs sampler) \citep{gamerman}.
Posterior means are used to obtain point estimates and standard deviations for describing variation of inferred parameters. 
The Heidelberger and Welch convergence diagnostic \citep{Heidelberger} and effective sample size diagnostic \citep{coda} are used to determine whether convergence has been achieved for all parameters. 

Computational times for convergence of our MCMC schemes (code is available at \url{https://github.com/jhncl/LNA.git}) can be compared using estimates for the minimum effective sample size per second (ESS\textsubscript{min}/sec). 
The average ESS\textsubscript{min}/sec of our approximate model (coded in C) is $\sim$100 and ``exact'' model $\sim$1 (coded in JAGS \citep{rjags} with 15 imputed states between time points, chosen to maximise ESS\textsubscript{min}/sec).
We find that our C code is typically twice as fast as the simple MCMC scheme used by JAGS, indicating that our inference is ${\sim}50\times$  faster than an ``exact'' approach.
A more efficient ``exact'' approach could speed up further, say by another factor of 5, but our approximate approach will at least be an order of magnitude faster.
We use a burn-in of 600,000 and a thinning of 4,000 to obtain a final posterior sample size of 1,000 for MCMC convergence of all our models.

\begin{figure}[h!]
  \centering
\includegraphics[width=14cm]{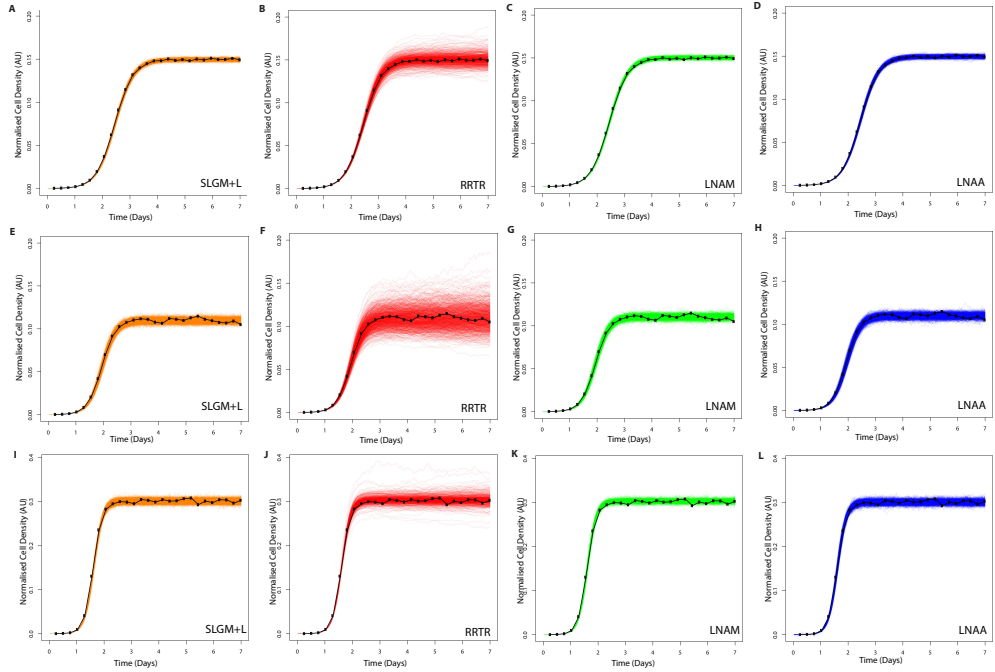}
\caption{\label{simlog}
Forward trajectories with measurement error, simulated from parameter posterior samples (sample size=1000). 
Model fitting is carried out on SLGM forward trajectories with lognormal measurement error (black), for three different sets of parameters (see Table~\ref{app:sde_val_fur}).  See (\ref{app:LNAM_sta_spa_mod}) or (\ref{app:LNAA_sta_spa_mod}) for model and Table~\ref{table:SDE_priors} for prior hyper-parameter values.
See Table~\ref{app:sde_val_fur} for parameter posterior means and true values.
A), E) \& I) SLGM+L (orange).
B), F) \& J) RRTR model with lognormal error (red).
C), G) \& K) LNAM model with lognormal error (green).
D), H) \& L) LNAA model with normal error (blue).
}
\end{figure}

To compare the approximate models ability to recover parameters from the SLGM with simulated lognormal measurement error, we simulate data and carry out Bayesian inference. Figure~\ref{simlog} shows that all three approximate models can capture the synthetic time-course well, but that the RRTR model is the least representative with the largest amount of drift occurring at the saturation stage, a property not found in the SLGM or the two new LNA models.
Comparing forwards trajectories with measurement error (Figure~\ref{simlog}), the ``exact'' model is visually similar to all our approximate models, but least similar to the RRTR.
Further, Table~\ref{app:sde_val_fur} demonstrates that parameter posterior means are close to the true values and that standard deviations are small for all models and each parameter set.
By comparing posterior means and standard deviations to the true values, Table~\ref{app:sde_val_fur} shows that all our models are able to recover the three different parameter sets considered. 

\begin{figure}[h!]
  \centering
\includegraphics[width=14cm]{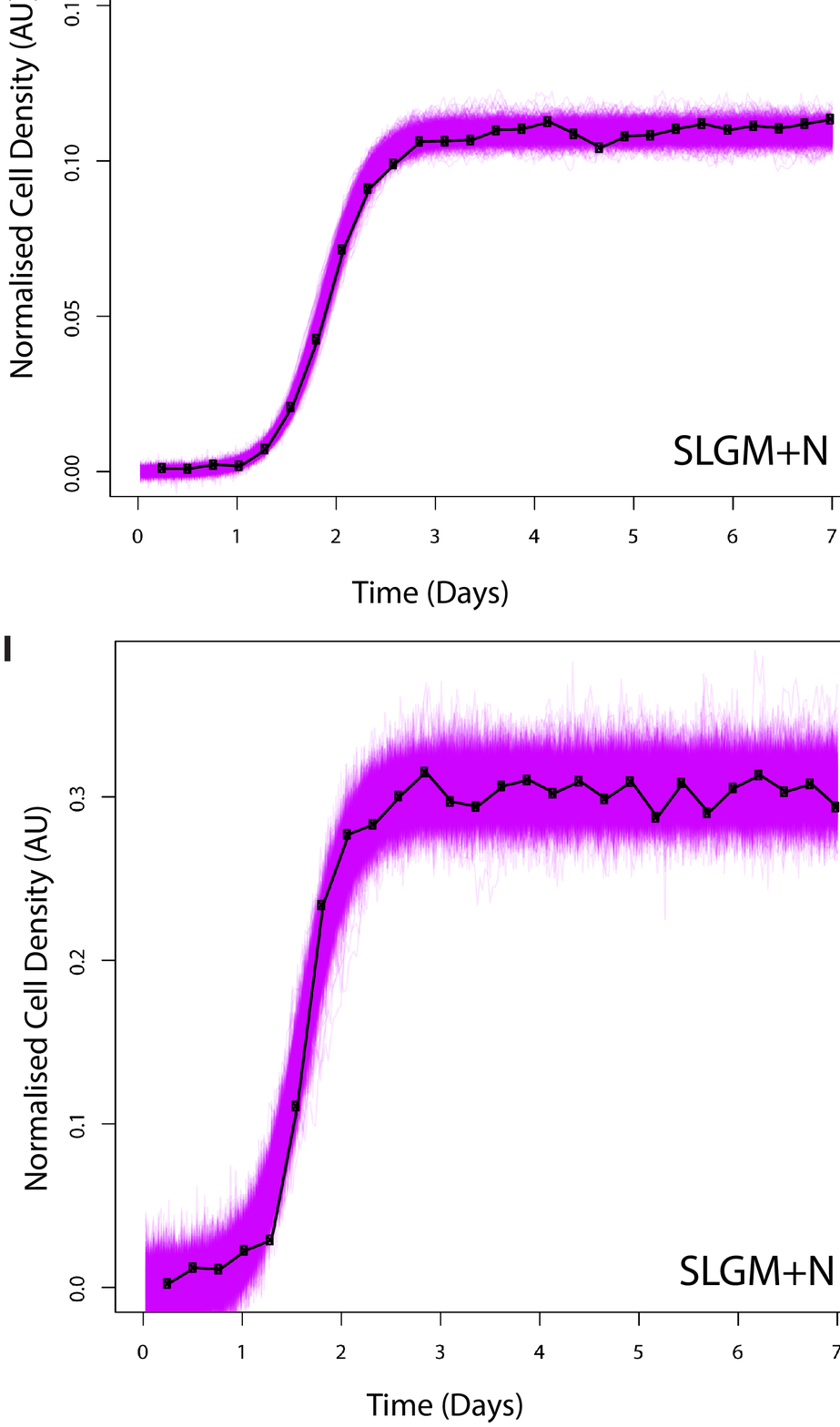}
\caption{\label{sim}
Forward trajectories with measurement error, simulated from inferred parameter posterior samples (sample size=1000).  
Model fitting is carried out on SLGM forward trajectories with normal measurement error (black), for three different sets of parameters (see Table~\ref{app:sde_val_fur}). 
See (\ref{app:LNAM_sta_spa_mod}) or (\ref{app:LNAA_sta_spa_mod}) for model and Table~\ref{table:SDE_priors} for prior hyper-parameter values.
See Table~\ref{app:sde_val_fur} for parameter posterior means.
A) SLGM+N (pink).
B) RRTR model with lognormal error (red).
C) LNAM model with lognormal error (green).
D) LNAA model with normal error (blue).
}
\end{figure}

To compare the approximations to the SLGM with simulated normal measurement error, we simulate data and carry out Bayesian inference. 
Figure~\ref{sim} shows that of our approximate models, only the LNAA model can appropriately represent the simulated time-course as both our models with lognormal measurement error, the RRTR and LNAM do not closely bound the data. 
Comparing forwards trajectories with measurement error (Figure~\ref{sim}), the ``exact'' model is most visually similar to the LNAA, which shares the same measurement error structure.
Further, Table~\ref{app:sde_val_fur} demonstrates that only our models with normal measurement error have posterior means close to the true values and that standard deviations are larger in the models with lognormal measurement error. 
Observing the posterior means for $K$ for each parameter set (Table~\ref{app:sde_val_fur}), we can see that the RRTR has the largest standard deviations and that, of the approximate models, its posterior means are furthest from both the true values and the ``exact'' model posterior means. 
Comparing LNA models to the ``exact'' models with matching measurement error, we can see in Table~\ref{app:sde_val_fur} that they share similar posterior means and only slightly larger standard deviations.
Example posterior diagnostics given in Appendix~\ref{app:diag_sim_n}, demonstrate that posteriors are distributed tightly around true values for our LNAA and data from the SLGM with Normal measurement error.

\begin{table}[h!]
\caption{\label{app:sde_val_fur}Bayesian state space model parameter posterior means, standard deviations and true values for Figure \ref{simlog}, \ref{sim} and \ref{real}. True values for the simulated data used for Figure \ref{4nonu}, \ref{simlog} and \ref{sim} are also given.}
\centering
\resizebox{\columnwidth}{!}{%
\npdecimalsign{.}
\nprounddigits{3}
\begin{tabular}{c c n{2}{5} n{1}{3} n{1}{3} n{1}{3} n{1}{3} n{1}{3} n{1}{3} n{1}{3} n{1}{3} n{1}{3}}
\hline
\emph{Panel} & \emph{Model} & \multicolumn{2}{c}{\emph{$\hat{K}$}}  & \multicolumn{2}{c}{\emph{$\hat{r}$}}  & \multicolumn{2}{c}{\emph{$\hat{P}$}} & \multicolumn{2}{c}{\emph{$\hat{\nu}$}} & \multicolumn{2}{c}{\emph{$\hat{\zeta}$}}  \\ 
\hline
\noalign{\vskip 0.4mm} 
\multicolumn{4}{l}{\emph{Figure \ref{simlog}, SLGM with lognormal error}}&&&&&&&&\\
\noalign{\vskip 0.2mm} 
A & SGLM+L & 0.1495369660 &( 0.001)&
2.9822427445 &(0.01350133)&
1.001567e-04 &(1.111598e-06)&
3.8597809e-03 &(2.126679e-03)&
0.0173228498 &(0.005037563)\\ 
B & RRTR & 0.149999335 &(0.00341)&
2.99015285 &(0.01092)&
9.93119e-05 &(1.06863e-06)&
5.6843212256e-03 &(2.36e-03)&
0.011547265557 &(0.00618)\\ 
C & LNAM & 0.149614787 &(0.001)&
2.98773783 &(0.01318)&
9.97960e-05 &(1.124219e-06)&
4.1396496436e-03 &(2.179554e-03)&
0.016364920984 &(0.00517)\\
D & LNAA & 0.14953488 &(0.0005)&
3.00514737 &(0.01995)&
9.64713e-05 &(2.945503e-06)&
3.098832e-05 &(2.53443e-05)&
0.0194997429 &(0.00344)\\ 
E & SGLM+L & 0.1096583 &(0.0007301528)&
3.974696 &(0.04724553)&
5.054189e-05 &(1.568196e-06)&
6.158735e-03 &(5.527275e-03)&
0.05052818 &(0.01393793)\\ 
F  & RRTR & 0.1086907   &(0.006533453)&
3.983695  &(0.03462259)&
5.046032e-05  &(1.137058e-06)&
5.928172e-03  &(4.596061e-03)&
0.03704644 &(0.00901516)\\ 
G  & LNAM & 0.1097371  &(0.0007490187)&
 3.985208  &(0.04606133)&
 5.043313e-05&(1.580017e-06)&
 6.187716e-03 &(5.190529e-03)&
  0.05156039 &(0.0126579)\\
H  & LNAA  &  0.1097912  &(0.0007867848)&
 3.959346  &(0.06704208)&
 5.207434e-05  &(4.309876e-06)&
 4.539877e-05   &(4.394698e-05)&
 0.05912248 &(0.00978808)\\ 
I
& SGLM+L & 0.3003784 &(0.0009748994)&
  5.996904  &(0.02851697)&
 1.962434e-05  &(4.0414e-07)&
 9.543053e-03 &(4.03492e-03)&
0.02406215 &(0.01543261)\\ 
J & RRTR & 0.3005997   &(0.0044928)&
6.015265  &(0.01697153)&
 1.942819e-05 &(2.835413e-07)&
 1.240917e-02 &(2.306528e-03)&
 0.007745952  &(0.006357571)\\ 
K & LNAM &  0.3004755 &(0.0009411441)&
  6.014510 &(0.03065974)&
1.953061e-05  &(4.202024e-07)&
8.942722e-03  &(4.251644e-03)&
0.02690633 &(0.01575695)\\
L & LNAA  &  0.3004586  &(0.001146215)&
 6.037006   &( 0.06747846)&
 1.895225e-05 &(1.502188e-06)&
  8.122229e-05  &(1.595971e-04)&
 0.04733552  &(0.007512705)\\ 
\noalign{\vskip 1mm} 
\multicolumn{4}{l}{\emph{Figure \ref{sim}, SLGM with normal error}}&&&&&&&&\\
\noalign{\vskip 0.2mm} 
A
& SLGM+N & 0.1502746 &(0.002206958)&
3.099127 &(0.08534504)&
9.298825e-05 &(7.304899e-06)&
5.326174e-03 &(1.008844e-03)&
0.05948689 &(0.02986999)\\ 
B & RRTR & 0.2127014398 &(0.12323)&
1.367943557 &(0.26339)&
4.55242737e-03 &(2.1184933e-03)&
2.5392518626e-01 &(1.0969e-01)&
0.41885683478 &(0.12943)\\ 
C  & LNAM & 0.1714519199 &(0.03280)&
1.579950397 &(0.27107)&
5.2409643e-03 &(2.0479954e-03)&
2.05421122203e-01 &(7.805e-02)&
0.4730700594  &(0.05100)\\ 
D & LNAA & 0.150452393 &(0.00222)&
2.9899791 &(0.26239)&
1.189e-04 &(7.098716e-05)&
5.49023564e-03 &(1.06e-03)&
0.052829431599  &(0.03326)\\
E
& SLGM+N &  0.1094272  &(0.0007586429)&
4.183277 &(0.07356718)&
 4.389757e-05 &(4.128667e-06)&
9.679088e-04  &(2.806267e-04)&
0.05662970 &(0.01165919)\\ 
F & RRTR & 0.1574587528 &(0.08749706)&
2.6313040010 &(0.3367076)&
4.398197e-04 &(1.677808e-04)&
1.040098786e-01 &(1.008875e-01)&
0.3735663206 &(0.1616813)\\ 
G  & LNAM & 0.1159608439 &(0.009422523)&
3.0185447100 &(0.373545)&
 4.967272e-04 &(1.396969e-04)&
3.34648648e-02 &(4.308955e-02)&
0.4753192595 &(0.04357369)\\ 
H & LNAA &  0.1095810  &(0.0007913119)&
4.009893 &( 0.1581109)&
5.011655e-05 &( 1.442577e-05)&
1.093103e-03 &(3.638245e-04)&
0.05275491 &(0.01322628)\\ 
I
& SLGM+N & 0.3052235554 &(0.00282216)&
5.2670916442 &(0.1249732)&
3.263387e-04 &(3.407446e-05)&
1.11938047e-02 &(1.974384e-03)&
0.0446024422 &(0.0310588)\\ 
J & RRTR & 0.313643513 &(0.05677519)&
3.029806490 &(0.2325609)&
1.307166e-03 &(2.897121e-04)&
2.22843445e-01 &(3.707941e-02)&
0.074919987 &(0.08628192)\\ 
K  & LNAM  & 0.312629722 &(0.02045301)&
3.391999960 &(0.4300475)&
1.118432e-03 &(3.2685e-04)&
1.17641652e-01 &(8.435387e-02)&
0.360004398 &(0.1647072)\\ 
L & LNAA &  0.3022532 &(0.002361439)&
 5.862218 &(0.5228873)&
2.889783e-05 &(2.598836e-05)&
8.773735e-03 &(1.466338e-03)&
0.04084204  &(0.02844386)\\
\noalign{\vskip 1mm} 
\multicolumn{4}{l}{\emph{ Figure \ref{real}, observed yeast data}}&&&&&&&&\\
\noalign{\vskip 0.2mm} 
A & SLGM+L &  0.1096916 &(0.007456264)&
4.098380 &(0.2994734)&
7.603410e-06 &(3.205816e-06)&
3.456518e-01 &(5.318567e-02)&
0.1128021 &(0.109331)\\ 
B & SLGM+N &  0.1098586 &(0.003273399)&
3.905401 &(0.1725181)&
1.043704e-05 &(3.085561e-06)&
1.851678e-04 &( 7.460388e-05)&
0.1666784 &(0.02814871)\\ 
C & RRTR &  0.1143094 &(0.02602507)&
3.763728 &(0.2006525)&
1.079002e-05 &(3.154911e-06)&
3.378525e-01 &(4.839998e-02)&
0.07771955 &(0.07732242)\\ 
D & LNAM & 0.1103702 &(0.01113706)&
3.776766 &(0.216272)&
1.077129e-05 &(3.277218e-06)&
3.36179e-01 &(5.137415e-02)&
0.1036413 &(0.1079923)\\
E & LNAA & 0.109176 &(0.003306315)&
3.832318 &(0.1984635)&
1.068951e-05 &(3.680477e-06)&
1.768693e-04 &(6.606697e-05)&
0.1637506 &(0.03259627)\\
\\ 
\hline
\multicolumn{2}{c}{\emph{True values}} & \multicolumn{2}{c}{\emph{K}}  & \multicolumn{2}{c}{\emph{r}}  & \multicolumn{2}{c}{\emph{P}} & \multicolumn{2}{c}{\emph{$\nu$}} & \multicolumn{2}{c}{\emph{$\sigma$}}   \\
\hline
\multicolumn{2}{c}{Figure~\ref{4nonu}, panels A, B, C and D} & \multicolumn{2}{c}{0.11} & \multicolumn{2}{c}{4} & \multicolumn{2}{c}{0.00005} & \multicolumn{2}{c}{N/A} & \multicolumn{2}{c}{0.05}  \\
\multicolumn{2}{c}{Figure~\ref{simlog} and \ref{sim}, panels A, B, C \& D}  & \multicolumn{2}{c}{0.15} & \multicolumn{2}{c}{3} & \multicolumn{2}{c}{0.0001} & \multicolumn{2}{c}{0.005} & \multicolumn{2}{c}{0.01} \\ 
\multicolumn{2}{c}{Figure~\ref{simlog} and \ref{sim}, panels E, F, G and H} & 
\multicolumn{2}{c}{0.11} & \multicolumn{2}{c}{4} & \multicolumn{2}{c}{0.00005} & \multicolumn{2}{c}{0.001} & \multicolumn{2}{c}{0.05}  \\
\multicolumn{2}{c}{Figure~\ref{simlog} and \ref{sim}, panels I, J, K and L} &
 \multicolumn{2}{c}{0.3} & \multicolumn{2}{c}{6} & \multicolumn{2}{c}{0.0002} & \multicolumn{2}{c}{0.01} & \multicolumn{2}{c}{0.02}  \\
\hline
\end{tabular}
\npnoround%
}
\end{table}

\subsection{\label{sec:application_obs}Application to observed yeast data}
\begin{figure}[h!]
  \centering
\includegraphics[width=14cm]{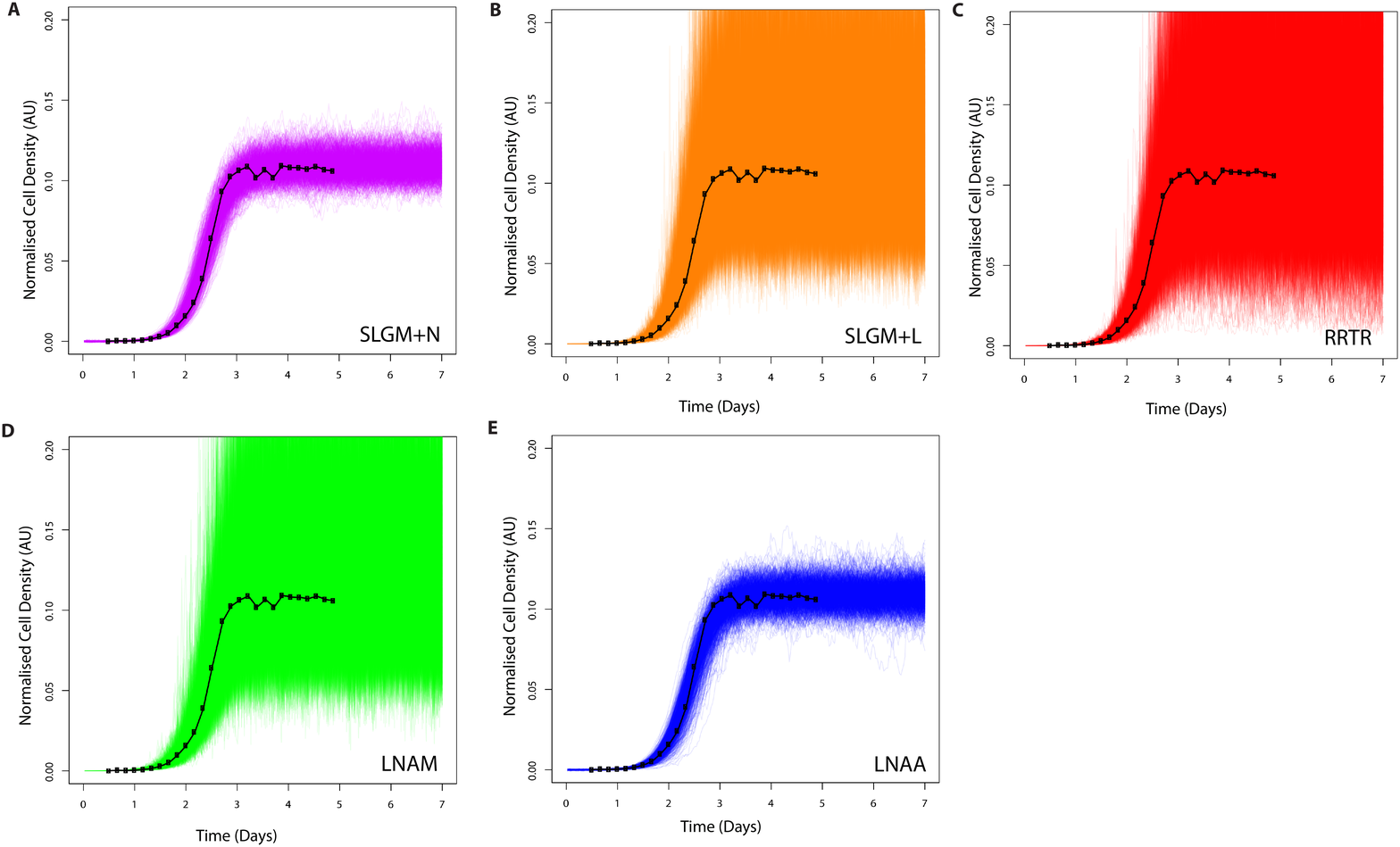}
\caption{\label{real}
Forward trajectories with measurement error, simulated from inferred parameter posterior samples (sample size=1000).  
Model fitting is carried out on observed yeast time-course data (black). 
See (\ref{app:LNAM_sta_spa_mod}) or (\ref{app:LNAA_sta_spa_mod}) and Table~\ref{table:SDE_priors} for prior hyper-parameter values.
See Table~\ref{app:sde_val_fur} for parameter posterior means.
A) SLGM+N (pink).
B) SLGM+L (orange).
A) RRTR model with lognormal error (red).
B) LNAM model with lognormal error (green).
C) LNAA model with normal error (blue).
}
\end{figure}

We now consider which diffusion equation model can best represent observed microbial population growth curves taken from a Quantitative Fitness Analysis (QFA) experiment \citep{QFA1,jove}, see Figure~\ref{real}. 
The data consists of scaled cell density estimates over time for budding yeast \emph{Saccharomyces cerevisiae}. 
Independent replicate cultures are inoculated on plates and photographed over a period of 5 days.
The images captured are then converted into estimates of integrated optical density (IOD, which we assume are proportional to cell population size), by the software package Colonyzer \citep{Colonyzer}.
The dataset chosen for our model fitting is a representative set of 10 time-courses, each with 27 time points.

As in Figure~\ref{sim}, we see that the LNAA model is the only approximation that can appropriately represent the time-course and that both the RRTR and LNAM fail to bound the data as tightly as the LNAA (Figure~\ref{real}).
Our two ``exact'' models are visually similar to our approximate models with the same measurement error, with the SLGM+N most similar to the LNAA and the SLGM+L to the RRTR and LNAM. This is as  expected due to matching measurement error structures.
Table~\ref{app:sde_val_fur} summarises parameter estimates for the observed yeast data using each model. The variation in the the LNAA model parameter posteriors is much smaller than the RRTR and LNAM, indicating a more appropriate model fit. 
Comparing the LNA models and ``exact'' models with matching measurement error, we can see in Table~\ref{app:sde_val_fur} that they share similar posterior means and standard deviations for all parameters and in particular, they are very similar for both $K$ and $r$, which are important phenotypes for calculating fitness \citep{QFA1}.

In Table~\ref{tab:MSE}, to compare quality of parameter inference for 10 observed yeast time-courses with each approximate model. Mean square error (MSE) for 1000 posterior sample forward simulations are calculated for each yeast time course and summed to give a Total MSE for each model.
It is clear that the RRTR is the worst overall representation of the 10 yeast time courses, with the highest total MSE and a much larger total MSE than the ``exact'' SLGM+L. It is interesting to see there is a very similar total MSE for the SLGM+L and LNAM, and similarly for the SLGM+N and LNAA, demonstrating that our approximations perform well.
\begin{table}[h!]
\caption{\label{tab:MSE} Total mean squared error (MSE) for 10 observed yeast growth time courses, each with 1000 forward simulated time-courses with measurement error. Parameter values are taken from posterior samples. Standard Deviations give the variation between the sub-total MSEs for each yeast time course fit (n=10). }
\centering
\npdecimalsign{.}
\nprounddigits{3}
\begin{tabular}{c n{1}{3} n{1}{3} n{1}{3} n{1}{3} n{1}{3}}
\hline
\emph{Model} & \multicolumn{1}{c}{{SLGM+N}}  & \multicolumn{1}{c}{{SLGM+L}}  &  \multicolumn{1}{c}{{RRTR}} & \multicolumn{1}{c}{{LNAM}} & \multicolumn{1}{c}{{LNAA}} \\ 
\hline
\noalign{\vskip 0.4mm} 
Total MSE & 29.84698 & 100.165 & 600.6007 & 99.39689 &  30.95872 \\
Standard Deviation & 1.688649 & 8.390729 &55.72009 &9.263094& 2.029686\\
\noalign{\vskip 0.4mm} 
\hline
\end{tabular}
\npnoround%
\end{table}

\clearpage
\section{\label{sec:conclusions}Conclusion}
We have presented two new diffusion processes for modelling logistic growth data where fast inference is required: the linear noise approximation (LNA) of the stochastic logistic growth model (SLGM) with multiplicative noise and the LNA of the SLGM with additive intrinsic noise. 
Both the LNAM and LNAA are derived from the linear noise approximation of the stochastic logistic growth model (SLGM).
The new diffusion processes approximate the SLGM more closely than an alternative approximation (RRTR) proposed by \cite{roman}.
The RRTR lacks a mean reverting property that is found in the SLGM, LNAM and LNAA, resulting in increasing variance during the stationary phase of population growth (see Figure~\ref{4nonu}).

We compared the ability of each of the three approximate models and the SLGM to recover parameter values from simulated datasets using standard MCMC techniques.  
When modelling stochastic logistic growth with lognormal measurement error we find that our approximate models are able to represent data simulated from the original process and that the RRTR is least representative, with large variation over the stationary phase (see Figure~\ref{simlog}).
When modelling stochastic logistic growth with normal measurement error we find that only our models with normal measurement error can appropriately bound data simulated from the original process (see Figure~\ref{sim}).
We also compared parameter posterior distribution summaries with parameter values used to generate simulated data after inference using both approximate and ``exact'' models (see Table~\ref{app:sde_val_fur}).  We find that, when using the RRTR model, posterior distributions for the carrying capacity parameter $K$ are less precise than for the LNAM and LNAA approximations.  We also note that it is not possible to model additive measurement error while maintaining a linear Gaussian structure (which allows fast inference with the Kalman filter) when carrying out inference with the RRTR.  We conclude that when measurement error is additive, the LNAA model is the most appropriate approximate model.  

To test model performance during inference with real population data, we fitted our approximate models and the ``exact'' SLGM to microbial population growth curves generated by quantitative fitness analysis (QFA) (see Figure~\ref{real}).
We found that the LNAA model was the most appropriate for modelling experimental data.  It seems likely that this is because a normal error structure best describes this particular dataset, placing the LNAM and RRTR models at a disadvantage.  We demonstrate that arbitrarily exact methods and our fast approximations perform similarly during inference for 10 diverse, experimentally observed, microbial population growth curves (see Table~\ref{tab:MSE}) which show that, in practise, our fast approximations are as good as ``exact'' methods.
We conclude that our LNA models are preferable to the RRTR for modelling QFA data. 

It is interesting to note that, although the LNAA is not a better approximation of the original SGLM process than the LNAM, it is still quite reasonable. Figure~\ref{4nonu}A~and~\ref{4nonu}D shows that the SLGM and LNAA processes are visually similar. Figure~\ref{4nonu}E demonstrates that forward trajectories of the LNAA also share similar levels of variation over time with the SLGM and LNAM.  

Fast inference with the LNAA gives us the potential to develop large hierarchical Bayesian models which simultaneously describe thousands of independent time-courses from QFA with a diffusion equation, allowing us to infer the existence of genetic interactions on a genome-wide scale using realistic computational resources. 

Here, we have concentrated on a biological model of population growth.  However, we expect that the approach we have demonstrated: generating linear noise approximations of stochastic processes to allow fast Bayesian inference with Kalman filtering for marginal likelihood computation, will be useful in a wide range of other applications where simulation is prohibitively slow.
 
\appendix
\appendix
\setcounter{figure}{0}
\renewcommand\thefigure{\thesection.\arabic{figure}}    
\setcounter{table}{0}
\renewcommand\thetable{\thesection.\arabic{table}}    

\section{\label{app:LNAM_sol} LNAM Solution}
\setcounter{figure}{0}
\setcounter{table}{0}
First we look to solve $dZ_t$, given in equation (\ref{eq:LNAM_dz}).
We define $f(t)=-\frac{baPe^{aT}}{bP(e^{aT}-1)+a}$ to obtain the following,
\begin{equation*}
dZ_t=f(t)Z_tdt+\sigma dW_t.
\end{equation*}
Define a new process $U_t=e^{-\int^t_{t_0}f(s)ds}Z_t$ and solve the integral,
\begin{equation*}
\int^t_{t_0}f(s)ds=\int^t_{t_0}-\frac{baPe^{aS}}{bP(e^{aS}-1)+a}ds=\log\left(\frac{a}{bP(e^{aT}-1)+a}\right),
\end{equation*}
where, $S=s-{t_0}$ and $T=t-{t_0}$.
Apply the chain rule to $U_t$,
\begin{equation*}
U_t=e^{-\int^t_{t_0}f(s)ds}dZ_t-f(t)e^{-\int^t_{t_0}f(s)ds}Z_tdt.
\end{equation*}
Now substitute in $dZ_t=f(t)Z_tdt+\sigma dW_t$ and simplify to give
\begin{equation*}
U_t= e^{-\int^t_{t_0}f(s)ds}\sigma dW_t.
\end{equation*}
Apply the following notation
$\phi(t)=e^{\int^t_{t_0}f(s)ds}=\frac{a}{bP(e^{aT}-1)+a}$ and $\psi(t)=\sigma$ to give
\begin{equation*}
U_t=\phi(t)^{-1}\psi(t) dW_t.
\end{equation*}
$U_t$, has the following solution,
\begin{equation*}
U_t=U_0+\int^t_{t_0} \phi(s)^{-1} \psi(s)dW_s.
\end{equation*}
As $U_t=\phi(t)^{-1}Z_t$, $Z_t$ then has the following solution,
\begin{equation*}
Z_t=\phi(t)\left[Z_0+\int^t_{t_0}\phi(s)^{-1}\psi(s) dW_s\right].
\end{equation*}
Finally, the distribution at time t is $Z_t|Z_0\sim N(M_t,E_t)$, where \\
$M_t=\phi(t)Z_0$ and 
$E_t=\phi(t)\int^t_{t_0}\left[{\phi(s)}^{-1}\psi(s)\right]\left[{\phi(s)}^{-1}\psi(s)\right]^{T}ds\:\phi(t)^T$.
\\
Further, $M_t=\frac{a}{bP(e^{aT}-1)+a}Z_0 $ and 
$E_t=\sigma^2\left[\frac{a}{bP(e^{aT}-1)+a}\right]^2
\int^t_{t_0}\left[
\frac{a}{bP(e^{aS}-1)+a}
\right]^{-2}$.\\
As 
$\int^t_{t_0}\left[
\frac{a}{bP(e^{aS}-1)+a}
\right]^{-2}ds=\frac{
b^2P^2e^{2aT}
+4bP(a-bP)e^{aT}
+2at(a-bP)^2
}{2a^3}
-
\frac{
b^2P^2
+4bP(a-bP)
+2at_0(a-bP)^2}{2a^3}
$,
\begin{align*}
E_t=&\sigma^2\left[\frac{a}{bP(e^{aT}-1)+a}\right]^2
\left[
\frac{
b^2P^2(e^{2aT}-1)
+4bP(a-bP)(e^{aT}-1)
+2aT(a-bP)^2
}{2a^3}
\right]\\
=&\sigma^2\left[\frac{
b^2P^2(e^{2aT}-1)
+4bP(a-bP)(e^{aT}-1)
+2aT(a-bP)^2
}
{2a\left(bP(e^{aT}-1)+a\right)^2}\right].
\end{align*}
Taking our solutions for $V_t$ (\ref{eq:LNAM_det_sol}) and $Z_t$, we can now write our solution for the LNA to the log of the logistic growth process (\ref{eq:SDE2}).\\
As $Y_t=V_t+Z_t$, 
\begin{equation*}
Y_t|Y_0\sim \mathcal{N}\left(\log\left[\frac{aPe^{aT}}{bP(e^{aT}-1)+a}\right]+M_t,E_t\right).
\end{equation*}
Note: $\frac{aPe^{aT}}{bP(e^{aT}-1)+a}$ has the same functional form as the solution to the deterministic part of the logistic growth process (\ref{eq_det_sde}) and is equivalent when $\sigma=0$ (such that $a=r-\frac{\sigma^2}{2}=r$).\\
\\
Further, as $Y_t$ is normally distributed, we know $X_t=e^{Y_t}$ will be log normally distributed and
\begin{equation*}
X_t|X_0\sim \log\:\mathcal{N}(\log\left(\frac{aPe^{aT}}{bP(e^{aT}-1)+a}\right)+M_t,E_t).
\end{equation*}
Alternatively set $Q=\left(\frac{\frac{a}{b}}{P}-1\right)e^{at_{0}}$,
\begin{equation*}
X_t|X_{0}\sim \log\:\mathcal{N}(\log\left(\frac{\frac{a}{b}}{1+Qe^{-at}}\right)+M_t,E_t).
\end{equation*}

\noindent From our solution to the log process we can obtain the following transition density
\begin{align*}
\begin{split}
(Y_{t_i}|Y_{t_{i-1}}&=y_{t_{i-1}})\sim\operatorname{N}\left(\mu_{t_i},\Xi_{t_i}\right),\\
\text{where } y_{t_{i-1}}&=v_{t_{i-1}}+z_{t_{i-1}}, Q=\left(\frac{\frac{a}{b}}{P}-1\right)e^{at_{0}},\\
\mu_{t_i}&=v_{t_{i-1}}+\log\left(\frac{1+Qe^{-at_{i-1}}}{1+Qe^{-at_i}}\right)+e^{-a(t_i-t_{i-1})}\frac{1+Qe^{-at_{i-1}}}{1+Qe^{-at_i}}z_{t_{i-1}} \text{ and}\\
\Xi_{t_i}&=\sigma^2\left[\frac{4Q(e^{at_i}-e^{at_{i-1}})+e^{2at_i}-e^{2at_{i-1}}+2aQ^2(t_i-t_{i-1})}{2a(Q+e^{at_i})^2}\right].
\end{split}
\end{align*}

\clearpage

\section{\label{app:LNAA_sol} LNAA Solution}
\setcounter{figure}{0}
\setcounter{table}{0}
First we look to solve $dZ_t$, given in (\ref{eq:LNAA_dz}).
We define $f(t)=a-2bV_t$ and $g(t)=aV_t-bV_t^2-\frac{a^2Pe^{aT}(a-bP)}{\left(bP(e^{aT}-1)+a\right)^2}$ to obtain the following,
\begin{equation*}
dZ_t=\left(g(t)+f(t)Z_t\right)dt+\sigma V_t dW_t.
\end{equation*}
Define a new process $U_t=e^{-\int^t_{t_0}f(s)ds}Z_t$ and solve the integral,
\begin{equation*}
\int^t_{t_0}f(s)ds=\int^t_{t_0}(a-2bV_s)ds=aT-2\log\left(\frac{bP(e^{aT}-1)+a}{a}\right),
\end{equation*}
as $\int^t_{t_0}V_sds=\frac{1}{b}\log \left(\frac{bP(e^{aT}-1)+a}{a}\right)$, where $S=s-{t_0}$ and $T=t-{t_0}$.
Apply the chain rule to $U_t$,
\begin{equation*}
U_t=e^{-\int^t_{t_0}f(s)ds}dZ_t-f(t)e^{-\int^t_{t_0}f(s)ds}Z_tdt.
\end{equation*}
Now substitute in $dZ_t=\left(g(t)+f(t)Z_t\right)dt+\sigma V_t dW_t$ and simplify to give,
\begin{equation*}
U_t=e^{-\int^t_{t_0}f(s)ds}g(t)dt+e^{-\int^t_{t_0}f(s)ds}\sigma V_t dW_t.
\end{equation*}
Apply the following notation $\phi(t)=e^{\int^t_{t_0}f(s)ds}=e^{aT}\left(\frac{a}{bP(e^{aT}-1)+a}\right)^2$ and $\psi(t)=\sigma V_t$ to give, 
\begin{equation*}
U_t=\phi(t)^{-1}g(t)+\phi(t)^{-1}\psi(t) dW_t.
\end{equation*}
$U_t$ has the following solution,
\begin{equation*}
U_t=U_0+\int^t_{t_0}\phi(s)^{-1}g(s)ds +\int^t_{t_0} \phi(s)^{-1} \psi(s)dW_s.
\end{equation*}
As $U_t=\phi(t)^{-1}Z_t$, $Z_t$ has the following solution,
\begin{equation*}
Z_t=\phi(t)\left[Z_0+\int^t_{t_0}\phi(s)^{-1}g(s)ds+\int^t_{t_0}\phi(s)^{-1}\psi(s) dW_s\right].
\end{equation*}
Finally the distribution at time t is $Z_t|Z_0\sim N(M_t,E_t)$, where \\
$M_t=\phi(t)\left(Z_0+\int^t_{t_0}\phi(s)^{-1}g(s)ds\right)$ and 
$E_t=\phi(t)\int^t_{t_0}\left[{\phi(t)}^{-1}\psi(t)\right]\left[{\phi(t)}^{-1}\psi(t)\right]^{T}ds\:\phi(t)^T$.
\begin{align*}
M_t&=\phi(t)\left(Z_0+ 
\int^t_{t_0}\left[\left(\phi(t)\right)^{-1} \left(aV_s-bV_s^2-\frac{a^2Pe^{aS}(a-bP)}{\left(bP(e^{aS}-1)+a\right)^2}\right)\right] ds
\right)  \\
&=\phi(t)\left(Z_0+ 
\int^t_{t_0}\left[\left(\phi(t)\right)^{-1} \left(0\right)\right] ds
\right) \\
&=e^{aT}\left(\frac{a}{bP(e^{aT}-1)+a}\right)^2Z_0
\end{align*}
 and 
\begin{equation*}
E_t=\left(e^{aT}\left(\frac{a}{bP(e^{aT}-1)+a}\right)^2\right)^2
\int^t_{t_0}\left[e^{aS}\left(\frac{a}{bP(e^{aS}-1)+a}\right)^2\right]^{-2} \sigma^2 V_s^2 ds
\end{equation*}
\begin{equation*}
=\sigma^2\left(e^{aT}\left(\frac{a}{bP(e^{aT}-1)+a}\right)^2\right)^2
\int^t_{t_0}\left[
e^{aS}\left(\frac{a}{bP(e^{aS}-1)+a}\right)^2
\right]^{-2} \left[
\frac{aPe^{aS}}{bP(e^{aS}-1)+a}
\right]^{2} ds.
\end{equation*}
\begin{equation*}
=\sigma^2\left(e^{aT}\left(\frac{a}{bP(e^{aT}-1)+a}\right)^2\right)^2
\int^t_{t_0}\left[
e^{-2aS}\left(\frac{a}{bP(e^{aS}-1)+a}\right)^{-4}
\right] \left[
\frac{aPe^{aS}}{bP(e^{aS}-1)+a}
\right]^{2} ds
\end{equation*}
\begin{equation*}
=\sigma^2\left(e^{aT}\left(\frac{1}{bP(e^{aT}-1)+a}\right)^2\right)^2
\int^t_{t_0}\left[a^2P^2
\left(\frac{1}{bP(e^{aS}-1)+a}\right)^{-2}
\right] ds,
\end{equation*}
as $\int^{t_{i}}_{t_{0}}\left(\frac{1}{bP(e^{aS}-1)+a}\right)^{-2}ds
=\frac{
b^2P^2e^{2aT}
+4bP(a-bP)e^{aT}
+2at(a-bP)^2
}{2a}
-
\frac{
b^2P^2
+4bP(a-bP)
+2at_0(a-bP)^2}{2a}$,
\begin{align*}
E_t=&\frac{1}{2}\sigma^2aP^2e^{2aT}\left(\frac{1}{bP(e^{aT}-1)+a}\right)^4\\
&\times
\left[
b^2P^2(e^{2aT}-1)
+4bP(a-bP)(e^{aT}-1)
+2aT(a-bP)^2
\right].
\end{align*}
\noindent From our solution to the process we can obtain the following transition density
\begin{align*}
\begin{split}
(X_{t_i}|X_{t_{i-1}}=&x_{t_{i-1}})\sim N(\mu_{t_i},\Xi_{t_i}),\\
\text{where }x_{t_{i-1}}=&v_{t_{i-1}}+z_{t_{i-1}},\\
\mu_{t_i}=&v_{t_{i-1}}+\left(\frac{aPe^{aT_i}}{bP(e^{aT_i}-1)+a}\right)-\left(\frac{aPe^{aT_{i-1}}}{bP(e^{aT_{i-1}}-1)+a}\right)\\
&+e^{a(t_i-t_{i-1})}\left(\frac{bP(e^{aT_{i-1}}-1)+a}{bP(e^{aT_i}-1)+a}\right)^2Z_{t_{i-1}}\text{ and}\\
 \Xi_t=&\frac{1}{2}\sigma^2aP^2e^{2aT_i}\left(\frac{1}{bP(e^{aT_i}-1)+a}\right)^4\\
&\times[
b^2P^2(e^{2aT_i}-e^{2aT_{i-1}})
+4bP(a-bP)(e^{aT_i}-e^{aT_{i-1}})\\
&\;\:\:\:\:+2a(t_i-t_{i-1})(a-bP)^2
].
\end{split}
\end{align*}
\clearpage

\section{\label{app:prior_hyp}Prior Hyper-parameters for Bayesian State Space Models}
\begin{table}[h!]
\caption{Prior hyper-parameters for Bayesian sate space models, Lognormal with mean ($\mu$) and precision ($\tau$) \label{table:SDE_priors}}
\centering     
\begin{tabular}{c c}
    \hline
		\noalign{\vskip 0.4mm} 
    Parameter Name  & \multicolumn{1}{c}{Value} \\ \hline
 ${\mu}_K$ & $\log(0.1)$   \\ 
 $\tau_{K}$ & 2 \\ 
 ${\mu}_r$ & $\log(3)$  \\ 
 $\tau_{r}$ & 5\\ 
 ${\mu}_P$ &  $\log(0.0001)$\\ 
 $\tau_{P}$ & 0.1 \\ 
 ${\mu}_\sigma$ & $\log(100)$  \\ 
 $\tau_{\sigma}$ & 0.1 \\ 
 ${\mu}_\nu$ & $\log(10000)$  \\ 
 $\tau_{\nu}$ & 0.1 \\ 
    \hline
    \end{tabular}
    \npnoround
\end{table}
\clearpage

	\clearpage

\section{\label{app:kalman_fil}Kalman Filter}
\setcounter{figure}{0}
\setcounter{table}{0}
To find $\pi(y_{t_{1:N}})$ for the LNAA with normal measurement error we can use the following Kalman Filter algorithm. First we assume the following:
\begin{align*}
\theta_{{t_{i}}}|y_{1:{{t_{i}}}}&\sim \operatorname{N}(m_{t_{i}},C_{t_{i}}),\\
m_{t_{i}}&=a_{t_{i}}+R_{t_{i}}F(F^{T}R_{{t_{i}}}F+U)^{-1}[y_{t_{i}}-F^{T}a_{t_{i}}],\\
C_{t_{i}}&=R_{t_{i}}-R_{t_{i}}F(F^TR_{t_{i}}F+U)^{-1}F^{T}R_{t_{i}}
\end{align*}
and initialize with $m_0=P$ and $C_0=0$. Now suppose that,
\begin{align*}
\theta_{t_{i}}|y_{1:{t_{i-1}}}&\sim \operatorname{N}(a_{t_{i}},R_{t_{i}}),\\
a_{t_{i}}&=G_{{t_{i}}}m_{{t_{i-1}}}\\  
\text{and }R_{t_{i}}&=G_{{t_{i}}}C_{{t_{i-1}}}G_{t_{i}}^T+W_{t_{i}}.
\end{align*}
The transition density distribution, see (\ref{eq:LNAA_tran}) is as follows:
\begin{align*}
\theta_{{t_{i}}}|\theta_{{t_{i-1}}}&\sim\operatorname{N}(G_{{t_{i}}}\theta_{{t_{i-1}}},W_{t_{i}})\\
\text{or equivalently }(X_{t_i}|X_{t_{i-1}}=x_{t_{i-1}})&\sim\operatorname{N}\left(\mu_{t_i},\Xi_{t_i}\right),\text{ where }x_{t_{i-1}}=v_{t_{i-1}}+z_{t_{i-1}},\\
\theta_{t}
 &=
\begin{pmatrix}
  1 \\
	X_{t_{i}}
 \end{pmatrix}
 =
 \begin{pmatrix}
  1 & 0 \\
  H_{\alpha,t_{i}} & H_{\beta,t_{i}}
 \end{pmatrix}
 \begin{pmatrix}
  1 \\
	X_{t_{i-1}}
 \end{pmatrix}
\\&=
 G_{t_{i}}\theta_{{t_{i-1}}},\\
 G_{t_{i}}&= \begin{pmatrix}
 1 & 0\\
 H_{\alpha,t_{i}} & H_{\beta,t_{i}}
 \end{pmatrix}, \quad
  W_{t_{i}}= \begin{pmatrix}
  0 & 0 \\
  0 & \Xi_{t_i}
	\end{pmatrix} \\
\text{where }H_{\alpha,t_{i}}=H_\alpha({t_{i}},{t_{i-1}})=&V_t-V_{t-1}e^{a(t_i-t_{i-1})}\left(\frac{bP(e^{aT_{i-1}}-1)+a}{bP(e^{aT_i}-1)+a}\right)^2\\
\text{and }H_{\beta,t_{i}}=&H_\beta({t_{i}},{t_{i-1}})=e^{a(t_i-t_{i-1})}\left(\frac{bP(e^{aT_{i-1}}-1)+a}{bP(e^{aT_i}-1)+a}\right)^2.
\end{align*}
The measurement error distribution is as follows:
\begin{align*}
y_{t_{i}}|\theta_{{t_{i}}}{\sim}&\operatorname{N}(F^T\theta_{{t_{i}}},U)\\
\text{or equivalently }y_{t_{i}}|\theta_{{t_{i}}}{\sim}&\operatorname{N}(X_{{t_{i}}},\sigma_{\nu}^2),\\
 \text{where }
 F=& \begin{pmatrix}
  0 \\
	1 
 \end{pmatrix}\text{ and }
  U=  \sigma_{\nu}^2.
  \end{align*}  
  Matrix Algebra:
\begin{align*}
a_{t_{i}}=&G_{{t_{i}}}m_{{t_{i-1}}}\\
=&\begin{pmatrix}
 1 & 0\\
H_{\alpha,t_{i}} & H_{\beta,t_{i}}
 \end{pmatrix}
 \begin{pmatrix}
  1\\
	m_{{t_{i-1}}} 
 \end{pmatrix}
=\begin{pmatrix}
  1\\
	H_{\alpha,t_{i}}+H_{\beta,t_{i}}m_{{t_{i-1}}}
 \end{pmatrix}
  \end{align*}  
  \begin{align*}
R_{t_{i}}&=G_{{t_{i}}}C_{{t_{i-1}}}G_{t_{i}}^T+W_{t_{i}}
\\
 &=
  \begin{pmatrix}
  0 & 0\\
  0 & {H_{\beta,t_{i}}}^2c_{{t_{i-1}}}^2
 \end{pmatrix}
 +
  \begin{pmatrix}
  0 & 0\\
  0 & \Xi_{t_i}
 \end{pmatrix}
   =\begin{pmatrix}
  0 & 0\\
  0 & {H_{\beta,t_{i}}}^2 c_{{t_{i-1}}}^2+\Xi_{t_i}
 \end{pmatrix}
\end{align*}
  \begin{align*}
  C_{t_{i-1}}&=
  \begin{pmatrix}
	0 & 0\\
  0 & c_{{t_{i-1}}}^2
 \end{pmatrix}
    \end{align*}
 \begin{align*}
  R_{t_{i}}F(F^{T}R_{{t_{i}}}F+U)^{-1}=&
  \begin{pmatrix}
	0 & 0\\
  0 & {H_{\beta,t_{i}}}^2c_{{t_{i-1}}}^2+\Xi_{t_i}
 \end{pmatrix}
 \begin{pmatrix}
  0 \\
	1
 \end{pmatrix}
\\
&\times
 \left[
 \begin{pmatrix}
 0 & 1\\
 \end{pmatrix}
   \begin{pmatrix}
	0 & 0\\
  0 & {H_{\beta,t_{i}}}^2c_{{t_{i-1}}}^2+\Xi_{t_i}
 \end{pmatrix}
  \begin{pmatrix}
  0 \\
	1
 \end{pmatrix}
 +\sigma_{\nu}^2
 \right]^{-1}
 \\
  =&\left[
    \begin{pmatrix}
{H_{\beta,t_{i}}}^2c_{{t_{i-1}}}^2+\Xi_{t_i} +\sigma_{\nu}^2
 \end{pmatrix}
 \right]^{-1}
 \begin{pmatrix}
	0 \\
  {H_{\beta,t_{i}}}^2c_{{t_{i-1}}}^2+\Xi_{t_i}
 \end{pmatrix}
\end{align*}
\begin{align*}
m_{t_{i}}=&a_{t_{i}}+R_{t_{i}}F(F^{T}R_{{t_{i}}}F+U)^{-1}[y_{t_{i}}-F^{T}a_{t_{i}}]\\
=&\begin{pmatrix}
  1\\
	H_{\alpha,t_{i}}+H_{\beta,t_{i}}m_{{t_{i-1}}}
 \end{pmatrix}\\
 &+
 \left[
    \begin{pmatrix}
{H_{\beta,t_{i}}}^2c_{{t_{i-1}}}^2+\Xi_{t_i} +\sigma_{\nu}^2
 \end{pmatrix}
 \right]^{-1}
 \begin{pmatrix}
	0 \\
  {H_{\beta,t_{i}}}^2c_{{t_{i-1}}}^2+\Xi_{t_i}
 \end{pmatrix}
 \left[
 y_{t_{i}}-
 \begin{pmatrix}
  0 & 1 \\
 \end{pmatrix}
 \begin{pmatrix}
  1\\
	H_{\alpha,t_{i}}+H_{\beta,t_{i}}m_{{t_{i-1}}}
 \end{pmatrix}
 \right]\\
   =&\begin{pmatrix}
	0\\
   H_{\alpha,t_{i}}+H_{\beta,t_{i}}m_{{t_{i-1}}}+\frac{{H_{\beta,t_{i}}}^2c_{{t_{i-1}}}^2+\Xi_{t_i}}{ {H_{\beta,t_{i}}}^2c_{{t_{i-1}}}^2+\Xi_{t_i}+\sigma_{\nu}^2}\left[y_{t_{i}}-H_{\alpha,t_{i}}-H_{\beta,t_{i}}m_{{t_{i-1}}}\right]
 \end{pmatrix}
 \end{align*}
   \begin{align*}
   C_{t_{i}}=&R_{t_{i}}-R_{t_{i}}F(F^TR_{t_{i}}F+U)^{-1}F^{T}R_{t_{i}}\\
 =&\begin{pmatrix}
	0 & 0\\
  0 & {H_{\beta,t_{i}}}^2c_{{t_{i-1}}}^2+\Xi_{t_i}
 \end{pmatrix}\\
&-
 \left[
    \begin{pmatrix}
{H_{\beta,t_{i}}}^2c_{{t_{i-1}}}^2+\Xi_{t_i} +\sigma_{\nu}^2
 \end{pmatrix}
 \right]^{-1}
 \begin{pmatrix}
	0\\
  {H_{\beta,t_{i}}}^2c_{{t_{i-1}}}^2+\Xi_{t_i}
 \end{pmatrix}
 \left[ 
  \begin{pmatrix}
 0 & 1
 \end{pmatrix}
  \begin{pmatrix}
  0 & 0\\
  0 & {H{\beta,t_{i}}}^2c_{{t_{i-1}}}^2+\Xi_{t_i}
 \end{pmatrix}
 \right]
 \\
     =&
       \begin{pmatrix}
  0 & 0 \\
  0 & {H_{\beta,t_{i}}}^2c_{{t_{i-1}}}^2+\Xi_{t_i}-\frac{\left({H_{\beta,t_{i}}}^2c_{{t_{i-1}}}^2+\Xi_{t_i}\right)^2}{{H_{\beta,t_{i}}}^2c_{{t_{i-1}}}^2+\Xi_{t_i}+\sigma_{\nu}^2}
 \end{pmatrix}
      \end{align*}
\\
With $m_{t_{i}}$ and $C_{t_{i}}$ for $i=1:T$, we can evaluate $a_{t_{i}}$, $R_{t_{i}}$ and $\pi(x_{t_{i}}|y_{t_{1:(i-1)}})$ for $i=1:T$.
We are interested in $\pi(y_{t_{1:i}})=\prod^T_{i=1}\pi(y_{t_{i}}|y_{t_{1:(i-1)}})$, where
$\pi(y_{t_{i}}|y_{t_{1:(i-1)}})=\int_{x}\pi(y_{t_{i}}|x_{t_{i}})\pi(x_{t_{i}}|y_{t_{1:(i-1)}})dx_{t_{i}}$ gives a tractable Gaussian integral.
Finally, 
\begin{align*}
\log\pi(y_{{t_{1:(i-1)}}})&=\sum^T_{i=1}\log\pi(y_{t_{i}}|y_{{t_{1:(i-1)}}})\\
&=\sum^T_{i=1}\left[-\log\left({\sqrt{2\pi(\sigma_{f}^2+\sigma_{g}^2)}}\right){-\frac{(\mu_f-\mu_g)^2}{2(\sigma_{f}^2+\sigma_{g}^2)}}\right],
\end{align*}
\begin{align*}
\text{where }\mu_f-\mu_g=&y_{t_{i}}-a_{t_{i}}=y_{t_{i}}-H_{\alpha,t_{i}}-H_{\beta,t_{i}}m_{t_{i-1}}\\
\text{and }\sigma_{f}^2+\sigma_{g}^2=&\sigma_{\nu}^2+R_{t_{i}}=\sigma_{\nu}^2+{H_{\beta,t_{i}}}^2c_{{t_{i-1}}}^2+\Xi_{t_i}.
\end{align*}
  \clearpage
\underline{Procedure}\\
\\
1. Set $i=1$. Initialize $m_0=\log(P)$ and $C_0=0$. \\
\\
2. Evaluate and store the following log likelihood term: 
\begin{align*}
\log\pi(y_{t_{i}}|y_{t_{1:(i-1)}})=&\left[-\log\left({\sqrt{2\pi(\sigma_{f}^2+\sigma_{g}^2)}}\right){-\frac{(\mu_f-\mu_g)^2}{2(\sigma_{f}^2+\sigma_{g}^2)}}\right],\\
\text{where }\mu_f-\mu_g=&y_{t_{i}}-H_{\alpha,t_{i}}-H_{\beta,t_{i}}m_{{t_{i-1}}}
\text{ and }\sigma_{f}^2+\sigma_{g}^2=\sigma_{\nu}^2+{H_{\beta,t_{i}}}^2c_{{t_{i-1}}}^2+\Xi_{t_i}.
\end{align*}
3. Create and store both $m_{t_i}$, and $C_{t_{i}}$,
\begin{align*}
 \text{where }m_{t_{i}}=&H_{\alpha,t_{i}}+H_{\beta,t_{i}}m_{t_{i-1}}+\frac{{H_{\beta,t_{i}}}^2c_{{t_{i-1}}}^2+\Xi_{t_i}}{ {H_{\beta,t_{i}}}^2c_{{t_{i-1}}}^2+\Xi_{t_i}+\sigma_{\nu}^2}\left[y_{t_{i}}-H_{\alpha,t_{i}}-H_{\beta,t_{i}}m_{{t_{i-1}}}\right]\\
\text{and }c_{{t_{i}}}^2=&{H_{\beta,t_{i}}}^2c_{{t_{i-1}}}^2+\Xi_{t_i}-\frac{\left({H_{\beta,t_{i}}}^2c_{{t_{i-1}}}^2+\Xi_{t_i}\right)^2}{{H_{\beta,t_{i}}}^2c_{{t_{i-1}}}^2+\Xi_{t_i}+\sigma_{\nu}^2}.
 \end{align*}
\\
4. Increment $i$, $i$=$(i+1)$ and repeat steps 2-3 till $\log\pi(y_{t_{N}}|y_{t_{1:(N-1)}})$ is evaluated.\\
\\
5. Calculate the sum:
\begin{equation*}
\log\pi(y_{{t_{1:N}}})=\sum^N_{i=1}\log\pi(y_{t_{i}}|y_{{t_{1:(i-1)}}}).
\end{equation*}

\clearpage

\section{\label{app:diag_sim_n}Parameter posterior means}
\setcounter{figure}{0}
\setcounter{table}{0}
\begin{figure}[h!]
  \centering
\includegraphics[width=14cm]{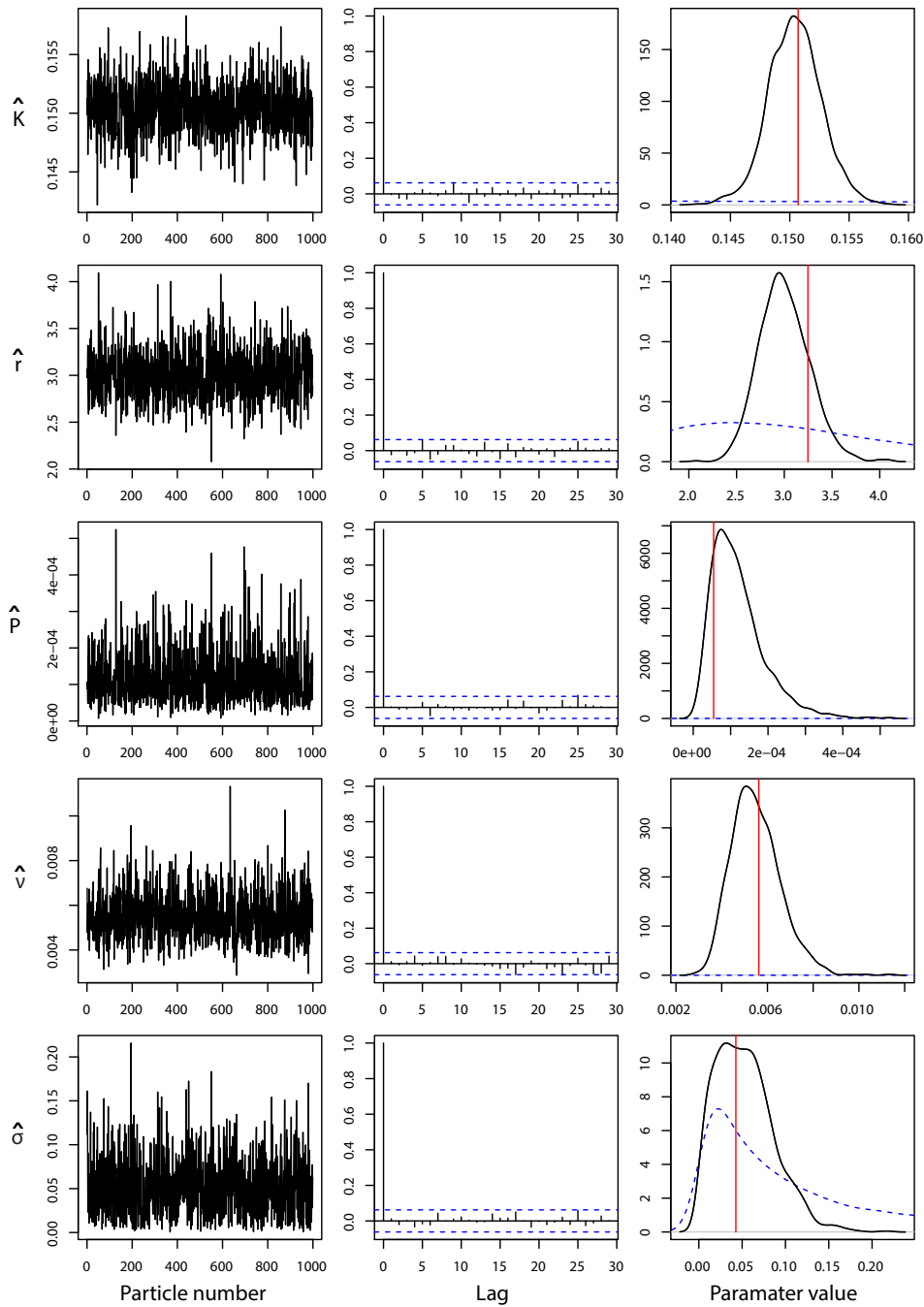}
\caption{Trace, auto-correlation and density plots for the LNAA model parameter posteriors (sample size = 1000, thinning interval = 4000), see Figure~\ref{sim}D.
Posterior density (black), prior density (dashed blue) and true parameter values (red) are shown in the right hand column.
}
\end{figure}

\section*{Acknowledgments}
This research was supported by grants from both the Biotechnology and Biological Sciences Research Council and the Medical Research Council.\vspace*{-8pt}
\bibliography{references} 
\bibliographystyle{Chicago}
\end{document}